\documentstyle[aps,prl,epsfig,balanced]{revtex}

\newcommand{\be}{\begin{equation}}
\newcommand{\ee}{\end{equation}}
\newcommand{\bc}{\begin{center}}
\newcommand{\ec}{\end{center}}
\newcommand{\bi}{\begin{itemize}}
\newcommand{\ei}{\end{itemize}}
\newcommand{\ba}{\begin{eqnarray}}
\newcommand{\ea}{\end{eqnarray}}
\newcommand{\ie}{{\it i.e.\ }}

\begin{document}
\draft
\twocolumn[\hsize\textwidth\columnwidth\hsize\csname
@twocolumnfalse\endcsname

\title{Growing Scale-Free Networks with Small World Behavior}
\author{Konstantin Klemm\cite{kk} and V\'{\i}ctor M. Egu\'{\i}luz\cite{vme}
\\Center for Chaos and Turbulence Studies\cite{cats}
\\ Niels Bohr Institute, Blegdamsvej 17, DK-2100 Copenhagen \O, Denmark}
\date{\today}
\maketitle
\begin{abstract}
In the context of growing networks, we introduce a simple dynamical model that
unifies the generic features of real networks: scale-free distribution of
degree and the small world effect. While the average shortest path length
increases logartihmically as in random networks, the clustering coefficient
assumes a large value independent of system size. We derive expressions for the
clustering coefficient in two limiting cases: random ($C \sim (\ln N)^2/N$)
and highly clustered ($C = 5/6$) scale-free networks.
\end{abstract}
]

Many systems can be represented by networks, \ie as a set of nodes joined
together by links indicating interaction. Social networks, the Internet,
food webs, distribution
networks, metabolic and protein networks, the networks of airline routes,
scientific collaboration networks and citation networks are just some examples
of such systems.
\cite{Strogatz01,Amaral00,Wasserman94,Watts98,Albert99,Williams00,Jeong00,Jeong01a,Redner98,Newman01a,Barabasi99}.
Most of these networks share three prominent features: (A) The average
shortest path length $L$ is small. In order to connect two nodes on the
graph, typically only a few edges need to be passed. (B) The clustering
coefficient $C$ is large. Two nodes having a common
neighbor are far more likely connected to each other than are two nodes
picked at random. (C) The distribution of the degree is scale-free, \ie
it decays as a power-law. The absence of a typical scale for the connectivity
of nodes is often related to the organization of the network as a
hierarchy.

In this Letter we present the first attempt to explain the
empirical observations by a model of network
self-organization according to simple rules. To our best
knowledge, all previous approaches of modeling complex networks have only
partially taken into account the above properties (A),(B) and (C).
Co-occurrence of high clustering and short distance between nodes was
originally termed as
the ``small world'' phenomenon. It can be obtained by departing from a
regular lattice, randomly rewiring links with a probability $p\ll1$
\cite{Watts98}. However, networks created in this way display a degree
distribution
sharply peaked around the mean-value; a power-law decay is not observed. 
Barab\'asi and Albert have given a first explanation of the scale-free
distribution by reformulating Simon's model
\cite{Simon55,Bornholdt00} in the context of growing networks.
New nodes join the network by attaching $m$
links to other nodes, chosen according to linear preferential attachment. This
means that a node obtains one of the new links with a probability proportional
to the number of links it already has. The algorithm, henceforth called BA
model, generates networks with a degree distribution $P(k)=2m^2k^{-3}$ with
$k\ge m$. However, as the system size $N$ grows, the clustering coefficient
approaches zero as the network size increases.
The value of the clustering coefficient predicted by the BA model is
typically several orders of magnitude lower than found empirically.

Recently an alternative algorithm has been suggested \cite{Klemm01} to account
for the high clustering found in scale-free networks. The topology of the
networks produced is similar to one-dimensional regular lattices. The
connectivity (coordination number), however, is not constant but follows a
power-law distribution causing the clustering to be even higher than in regular
lattices. Here we generalize the model to include long-range connections. We
find that a small ratio of long-range connections is sufficient to obtain small
path length, keeping the high clustering and scale-free degree distribution of
the original model.

Let us recall the high clustering model as originally defined in
Ref.~\cite{Klemm01}: Each node of the network is assigned a binary state
variable. A newly generated node is in the {\em active} state and keeps
attaching links until eventually deactivated. Taking a completely connected
network of $m$ active nodes as an initial condition, each step of the
time-discrete dynamics consists of the following three stages: (i) A new node
joins the network by attaching a link to each of the $m$ active nodes. (ii) The
new node becomes active. (iii) One of the active nodes is deactivated. The
probability that node $i$ is chosen for deactivation is $p_i = a k_i^{-1}$ with
normalization $a=\sum_j k_j^{-1}$. The model generates networks with degree
distribution $P(k)= 2 m^2 k^{-3}$ ($k\ge m$) and average connectivity $\langle
k \rangle = 2 m$ \cite{Klemm01}. Regarding topological properties the networks
are reminiscent of one-dimensional regular lattices. The path length increases
linearly with system size whereas the clustering coefficient quickly converges
to a constant value. 

\begin{figure}
\centerline{\epsfig{file=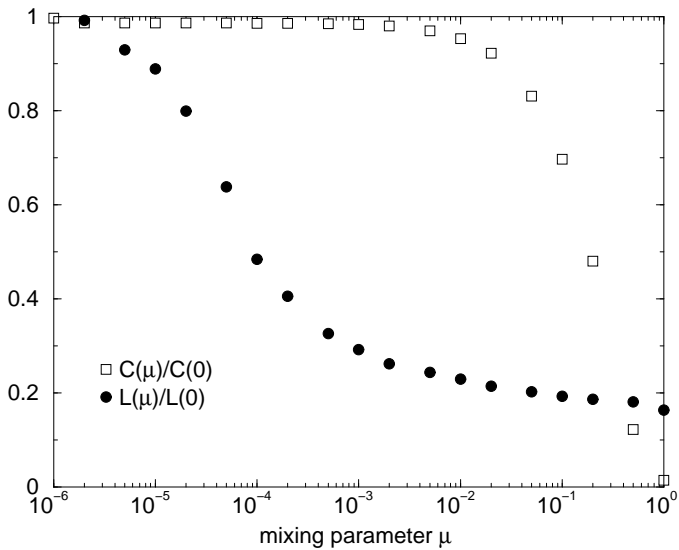,width=.5\textwidth}}
\caption{\label{fig1}Small world effect in scale-free networks. Introducing
ratio $\mu\ll1$ of random links into the highly clustered scale free
networks dratically reduces the typical distance $L$ between nodes. However
the strongly interconnected neighborhoods of the original model ($\mu=0)$
are preserved, as the clustering coefficient remains at its large value.
Only when $\mu$ reaches the order of $1$ the clustering
coefficient drops significantly. All plotted values are
averages over 100 independent realizations.
The networks have $N=10^4$ nodes with average degree $\langle k \rangle =20$.
} 
\end{figure}

Long-range connections are introduced into the model by modifying stage (i) in
the dynamical rules as follows. For each of the $m$ links of the new node it is
decided randomly whether the link connects to the active node (as in the
original model) or it connects to a random node. The latter case occurs with a
probability $\mu$. In this case the random node is chosen according to linear
preferential attachment, \ie the probability that node $j$ obtains a link is
proportional to the node's degree $k_j$. For $\mu = 0$ we recover the high
clustering model. The case $\mu = 1$ is the BA model. Varying $\mu$ in the
interval $[0,1]$ allows us to study the cross-over between the two models. We
are especially interested in the behaviour of the topological properties,
namely the average shortest path length and the clustering coefficient, as a
function of the cross-over parameter $\mu$.  Figure~\ref{fig1} shows the
variation of the average shortest path length and the clustering coefficient
with the parameter $\mu$. When increasing $\mu$ from zero to small finite values,
the average shortest path length $L$ drops rapidly and approaches the low value
of the BA model. The clustering coefficient $C$ remains practically constant in
this same range $0<\mu\ll1$. We have checked that the power law distribution of
the degree (not shown here) is still obtained in this range. Thus the model
with $0<\mu\ll1$ reproduces the three generic properties (A), (B) and (C) of
real-world networks. The model is robust against changes in the rule for the
introduction of random links. The small world transition shown in Fig.
\ref{fig1} does not change significantly when the attachment is not
preferential, i.e. every node receives a random link with the same probability.

\begin{figure}
\centerline{\epsfig{file=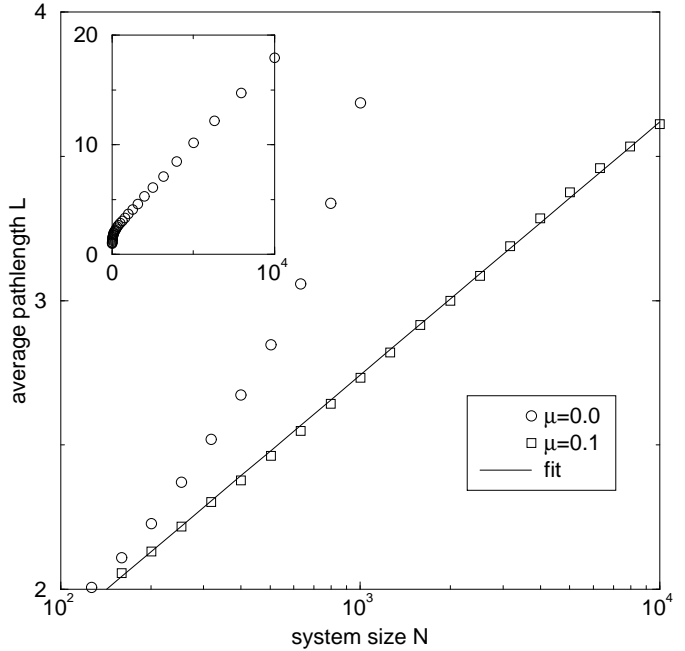,width=.5\textwidth}}
\caption{\label{fig2}Average shortest path length $L$ as a function of
system size $N$. In networks without long-range connections ($\mu=0$)
the relation between $L$ and $N$ is linear. This is seen best in the inset
with linear scales on both axes. When attaching a fraction $\mu=0.1$ of all
links to random nodes instead of the currently active ones, $L$ grows merely
logarithmically with $N$. The values can be fit well by a straight line in the
plot with logarithmic $N$ scale (main panel). All values plotted are
averages over 100 independent realizations. The average degree is $\langle k
\rangle=20$.}
\end{figure}

The observed drop in the average shortest path length, $L$, is due to a
qualitative change in the dependence of $L$ on the system size. In
Fig.~\ref{fig2} we show $L$ as a function of the system size $N$ for $\mu=0$
and $\mu=0.1$. For $\mu=0$, the average shortest path length grows linearly
$L\propto N$, the same behavior observed in one-dimensional regular lattices.
In clear contrast, a logarithmic growth of $L$ is obtained for $\mu=0.1$, $L
\propto \ln N$. The logarithmic increase of $L$ with system size is typical of
the small world effect \cite{Newman00}.

In the remainder of the Letter we study the evolution of the clustering
coefficient $C$ as a function of network size $N$. We begin by deriving
$C$ analytically for the two limiting cases $\mu=0$ (the high clustering
model) and $\mu=1$ (the BA model).

\begin{figure}
\centerline{\epsfig{file=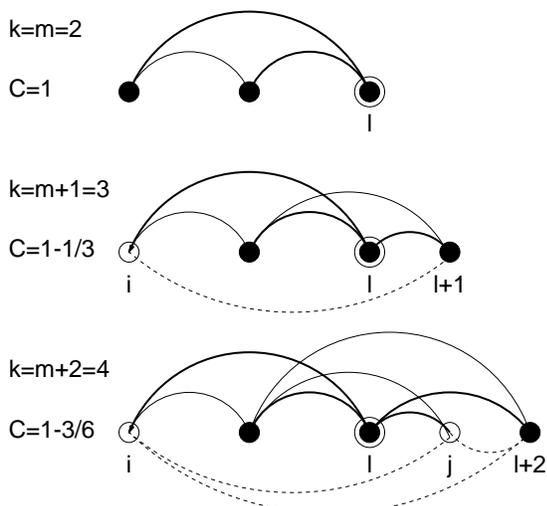,width=.4\textwidth}}
\caption{\label{fig3}Illustrating the calculation of the clustering
coefficient of the highly clustered model ($\mu=0$, $m=2$). The encircled
node is the node $l$ under consideration. Links of this node are drawn as
thick lines, links between its neighbors are thin lines. The dotted lines
are links that are ``missing'' in the neighborhood of $l$. Active nodes are
filled circles, inactive nodes are unfilled. Further explanation see text.
}
\end{figure}

Consider first the case $\mu=0$.
At any given time step the set of active nodes is completely interconnected,
simply because a newly generated node always connects to all active nodes
before being activated itself. It follows that a node $l$ with
degree $k_l=m$ has $C_l=1$ because all the $m(m-1)/2$ possible
links between neighbors of $l$ actually exist. If $l$ is deactivated in the
time step of its generation its neighborhood does not change any
more and it keeps $C_l=1$. Otherwise a node $i\neq l$ is deactivated. In the
next time step the node $l+1$ connects to $l$ and all its neighbours apart
from node $i$. Then $k_l(k_l-1) - 1$ of the possible $k_l(k_l-1)$ links between
neighbours of $l$ exist, where now $k_l=m+1$. If node $l$ keeps being active
a node $j\neq l$ is deactivated. Node $l+2$ connects to all neighbors of $l$
apart from $i$ and $j$ causing another 2 links to be missing in the
neighborhood of $l$. See Fig.~\ref{fig3} for an illustration. By induction
follows that after $n$ iterations $\sum_{\nu=1}^n \nu = n(n+1)/2$ links are
missing in the neighborhood of $l$.

Thus the clustering $C_l$ depends only on the degree $k_l$. The exact
relation is  
\be
C(k) = 1 - \frac{(k-m+1)(k-m)}{k(k-1)}~. 
\ee
The clustering coefficient $C$ can be obtained as the mean value of
$C(k)$ with respect to the degree distribution $P(k)=2m^2k^{-3}$, $k\geq m$. The
result is
\ba
C&=&\int_m^\infty  \left(1 - \frac{(k-m+1)(k-m)}{k(k-1)} \right)2m^2k^{-3} dk	\\
&=& \frac{5}{6} - \frac{7}{30m} + {\cal O} (m^{-2})~.
\ea 
In the limit of large $m$ the clustering coefficient is $5/6$. It is worth
noting that this value is higher than for regular lattices. The value
$5/6 \approx 0.83$ is similar to the one obtained in the film actor network
(0.79), the coauthorship network in neuroscience (0.76), and networks of word
synonyms (0.7) \cite{Albert01}.

Let us now consider the BA model ($\mu=1$). When adding node $j$ to the
network, the probability for one link of node $j$ to connect with node $i$ 
is the ratio of the degree of the node $i$, $k_i$, and the sum of all nodes'
degrees in the network, $2mj$. Thus the probability for the
existence of a link from $j$ to $i$ is given by
\be
\Pr\{(ij)\} = m \frac{k_i(j)}{2mj} ~,
\label{eqABpr1}
\ee
where the prefactor $m$ takes into account, that $m$ links per node are
added to the network. By $k_i(j)$ we denote the degree of node $i$ at the time
that node $j$ is added. Neglecting small fluctuations, the degree of the
$i$-th node is
$k_i(j)=m(j/i)^{0.5}$ according to Ref.~\cite{Barabasi99}. Inserting into Eq.\
(\ref{eqABpr1}) gives
\be
\Pr\{(ij)\} = \frac{m}{2}(ij)^{-0.5}~.
\label{eqABpr2}
\ee
The local clustering $C_l(N)$ of the node $l$ in a network of size $N$   
is defined as the number of links between neighbours of $l$, divided
by the total number of pairs of neighbors $l$ has. Only taking into
account expectation values and treating the nodes as a continuum,
we find
\be
C_l(N) = \frac{\int_1^N di \int_1^N dj \Pr\{(li)\} \Pr\{(lj)\} \Pr\{(ij)\}}
              {k_l^2(N)}~,
\ee
where we have approximated the total number of neighbors by $k_l^2/2$.
Evaluating the probabilities according to Eq.~(\ref{eqABpr1}) and using
$k_l^2(N) = m^2 N/l$ yields
\ba
C_l(N) &=& \frac{m^3}{8 k_l^2(N)} \int_1^N di \int_1^N dj\,(li)^{-0.5}(lj)^{-0.5}(ij)^{-0.5}\\
       &=& \frac{m^3}{8 lk_l^2(N)} (\ln N)^2\\
       &=& \frac{m}{8} \frac{(\ln N)^2}{N}~. \label{eqBAfinalC}
\ea
The average value of the local clustering $C_l$ does not
depend on the node $l$ under consideration. The networks generated by the
BA-model show homogeneous
clustering, despite the inhomogenous scale-free connectivity. With increasing
network
size $N$, the clustering coefficient decreases as $N^{-1}$ in leading order.
The difference with respect to a random graph, having a Poisson distribution
of degree, is seen only in the logarithmic correction $(\ln N)^2$.

\begin{figure}
\centerline{\epsfig{file=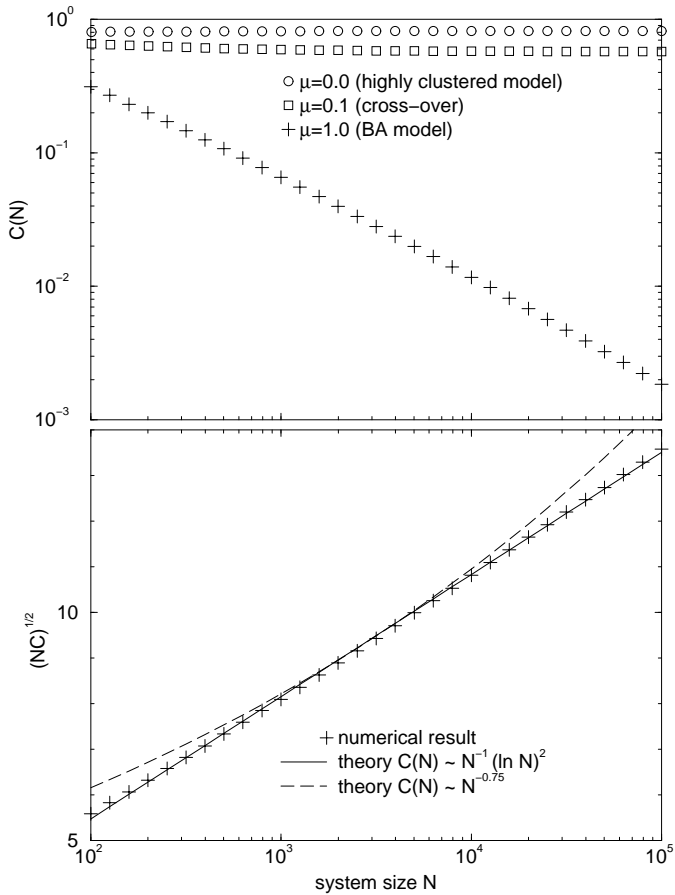,width=.5\textwidth}}
\caption{\label{fig4}Upper panel: The clustering coefficient $C$ as a
function of network size. Networks generated with $\mu=0.0$ quickly reach the
large value predicted by the analytical calculations ($C\approx 0.83$).
With 10\% long-range connections ($\mu=0.1$) the clustering is lower but
still approaches an asymptotic value
clearly above zero. In the BA-model ($\mu=1.0$) the clustering coefficient
decreases drastically with growing system size. Each of the three data sets
is an average over 100 independent simulation runs. 
Lower panel: For the BA model, the function
$(C(N)/N)^{0.5}$ grows as $\ln N$, giving a straight line in logarithmic-linear
plot. This indicates very good agreement with the analytical result
$C(N) \propto N^{-1}(\ln N)^2$. For comparison, the theoretical curve
$C(N) \propto N^{-0.75}$ is shown, as suggested in Ref.~\protect\cite{Albert01}.
}
\end{figure}

Figure \ref{fig4}, upper panel, shows the clustering coefficient obtained
from numerical simulations. For $\mu=0$ we find an asymptotic value
of approximately $0.83$ as predicted analytically. Also for $\mu=0.1$
convergence to a finite value is observed.
The BA model ($\mu=1.0$) displays a rapid decay of $C$ as the network size
$N$ grows. The behavior of $C(N)$ in the BA model is analyzed in the lower
panel of Fig.\ \ref{fig4}, clearly supporting the expression in Eq.\
\ref{eqBAfinalC}.
$C(N)$ is found to be inversly proportional to the system size, with
logarithmic corrections. A pure power law with exponent -0.75 as proposed in
Ref.~\cite{Albert01} describes the numerical data less accurately.

In summary, we have defined and analyzed a model of self-organizing networks
with high clustering, small path length and a scale-free distribution of
degree. The networks with these generic properties are obtained as a cross-over
between highly clustered scale-free networks \cite{Klemm01} and scale-free
random graphs \cite{Barabasi99}. The dependence of the topology on the
cross-over parameter is very similar to the small world transition observed
when introducing random links into a regular grid \cite{Watts98}. Therefore our
studies make a connection between small world graphs and scale-free networks,
essentially unifying both concepts in one model. 



\end{document}